\documentclass[prb,twocolumn,showpacs,preprintnumbers,amsmath,amssymb]{revtex4}

\usepackage{epsfig}
\usepackage{amsmath}
\usepackage{amsfonts}
\usepackage{bm}

\newcommand{\R}{{\bf R}}
\renewcommand{\l}{{\bf l}}
\renewcommand{\d}{{\bm{\delta}}}
\newcommand{\A}{{\bf A}}

\renewcommand{\j}{{\bf j}}
\renewcommand{\k}{{\bf k}}

\newcommand{\q}{{\bf q}}
\renewcommand{\r}{{\bf r}}

\newcommand{\ii}{{\rm i}}

\def\gl{\lower.35em\hbox{$\stackrel{\textstyle>}{\textstyle<}$}}

\begin{document}
\title{Dynamical current-current correlation of the hexagonal lattice and graphene}
\author{T. Stauber$^{1,2}$ and G. G\'omez-Santos$^1$}
\affiliation{$^1$Departamento de F\'{\i}sica de la Materia Condensada  and Instituto Nicol\'as Cabrera, Universidad Aut\'onoma de Madrid, E-28049 Madrid, Spain}
\affiliation{$^2$Centro de F\'{\i}sica  e  Departamento de
F\'{\i}sica, Universidade do Minho, P-4710-057, Braga, Portugal}
\date{\today}
\begin{abstract}
We discuss the dynamical current-current correlation function of the hexagonal lattice using a local current operator defined on a continuum-replica model of the original lattice model. In the Dirac approximation, the correlation function can be decomposed into a parallel and perpendicular contribution. We show that this is not possible for the hexagonal lattice even in the Dirac regime. A comparison between the analytical isotropic solution and the numerical results for the honeycomb lattice is given.  
\end{abstract}
\maketitle
%%%%%%%%%%%%%%%%%%%%%%%%%%%%%%%%%%%%%%%%%%%%%%%%%%%%%%%%%%%%%%%%%%%
% Section : Introduction
%%%%%%%%%%%%%%%%%%%%%%%%%%%
\section{Introduction}
Graphene is a two-dimensional carbon allotrope which was isolated in 2004\cite{Nov04} and has attracted immense research activities due to its novel mechanical and electronic properties.\cite{Geim09,Neto09,DasSarma10,Peres10} Whereas the mechanical properties are determined by electrons with $sp^2$-hybridization, the electronic properties can be mainly deduced considering only the $\pi$-electrons. The simplest model to study the electronic response of graphene to an external field or potential is thus given by a one-orbital tight-binding model on a hexagonal lattice.

Most of the novel electronic properties of graphene originate from the fact that there are two equivalent atoms in the Wigner-Seitz cell which give rise to two gapless bands with linear density of states close to the neutrality point. Most standard results of solid state text books can thus not be applied to the case of graphene due to the different dispersion and/or dimensionality, but also due to the two coherently coupled bands. 

An example is the density-density correlation or Lindhard function which in the case of the honeycomb lattice is given by\cite{Adler62} 
\begin{align}
\label{densitycorrelation}
\pi^{0,0}(\q,\omega)&=\frac{-g_s}{(2\pi)^2}\int_{\text{1.BZ}}d^2k\sum_{s,s'=\pm}f_{s\cdot s'}^{0,0}(\k,\q)\\\nonumber
&\times\frac{n_F(E^{s}({\bf k}))-n_F(E^{s'}({\bf k}+{\bf q}))}{E^{s}({\bf k})-E^{s'}({\bf k}+{\bf q})+\hbar\omega+\ii\delta}\;,
\end{align}
with the eigenenergies $E^{\pm}({\bf k})=\pm t|\phi(\k)|$ ($t\approx2.7$eV is the hopping amplitude), $n_F(E)$ the Fermi function, $g_s=2$ the spin-degeneracy and $\phi(\k)$ the complex structure factor defined below. Due to the two gapless bands, the above expression contains the band-overlap function 
\begin{align}
\label{overlapdensity}
f_{\pm}^{0,0}(\k,\q)&=\frac{1}{2}\left(1\pm\text{Re}\left[\frac{\phi(\k)}{|\phi(\k)|}\frac{\phi^*(\k+\q)}{|\phi(\k+\q)|}\right]\right)\;,
\end{align}
which marks the crucial difference to the standard text-book results containing only one band.\cite{footnote}

In the linear (Dirac) approximation of the band dispersion, the above expression can be solved analytically for finite chemical potential $\mu$ at zero temperature.\cite{Wunsch06,Hwang07} In the static case, it shows differences to the one-band result by Stern\cite{Stern67} for $|\q|\geq2k_F$ with $k_F$ the Fermi wave vector due to the contribution of  interband processes. For finite frequencies, these differences are even more pronounced and lead to a logarithmic singularity at $\hbar\omega=2\mu$.

The density-density correlation function or polarizability of graphene was calculated in a number of papers using different formalisms and introducing various modifications to the original Dirac Hamiltonian.\cite{Shung86,Gon94,Ando06,Vafek06,Mishchenko08,Ramezanali09,Roldan09,Pyatkovskiy09,Pedersen09,Ziegler09,Stauber10,Tudorovskiy09,Katsnelson10} Using these results, plasmons,\cite{Wunsch06,Hwang07,Ziegler09,Haas10} wrinkles,\cite{Gazit09} van-der-Waals interactions\cite{GomezSantos09} and forces due to moving external charges\cite{Allison09} were discussed. In this paper, we will focus on the related current-current correlation function $\pi^{i,j}(\q,\omega)$ of graphene ($i,j=x,y$) starting from the tight-binding model of the honeycomb lattice. 

In the Dirac approximation, the system is rotationally invariant and the current-current correlation function can be decomposed in a parallel $\pi^\parallel$ and perpendicular contribution $\pi^\perp$. The parallel contribution is related to the density-density correlation function via the continuity equation and thus determines the dielectric properties of the system. The perpendicular contribution is related to the magnetic susceptibility which in the static case has been first discussed by McClure\cite{McClure56} via the Helmholtz free energy and recently by Ando and co-workers using $\pi^\perp$.\cite{Ando09} In view of new experiments on the magnetic behavior of graphene\cite{Geim10}, the magnetic susceptibility was also calculated including electron-electron interactions to first order which results in a paramagnetic response away from half-filling.\cite{Polini10}

Here, we shall mainly discuss $\pi^\perp(\q,\omega)$ for finite frequencies. In the Dirac approximation, this was first done in Ref. \cite{Polini09}. We will summarize their results and compare the analytical solution of the isotropic system with the numerical solution of the hexagonal lattice. For that, we will define a local current operator defined for a continuous-replica model of the original lattice Hamiltonian. This formalism permits deeper insight in the lattice effects and can be used to calculate corrections which are lost in the scaling limit, i.e., the Dirac model.

The paper is organized as follows. In section II, we will define the continuum model and derive the local current operator of this model. We will further show that this operator satisfies the continuity equation with respect to the density operator defined on the lattice. In section III, we will present general expressions for the current-current correlation function and introduce the parallel and perpendicular contribution defined for the Dirac model. In section IV, we summarize the analytical results and compare them with the numerical results obtained from the hexagonal lattice. We close with a summary and conclusions and give real expressions for the current-current correlation function in an appendix.          

%%%%%%%%%%%%%%%%%%%%%%%%%%%%%%%%%%%%%%%%%%%%%%%%%%%%%%%%%%%%%%%%%%%
% Section : Continuum model and current operator
%%%%%%%%%%%%%%%%%%%%%%%%%%%
\section{Continuum model and current operator}
To calculate the current-current correlation function for a lattice model for finite wave vector $\q$, we are confronted with the following problem. The current operator for a lattice model, as given by the continuity equation, describes the flow from site $i$ to $j$ per unit time.\cite{Wen} In order to define a vector which depends on one lattice site instead of two, one needs to define a continuous model based on the Hamiltonian in reciprocal space. If the vector potential $\A$ is a smooth function of $\r$, then the coupling between $\A$ and the current can only see the smooth part and the continuous limit is justified.

Let us start with the tight-binding Hamiltonian of a general bipartite lattice with $N_c$ lattice sites $\R$ and nearest-neighbor lattice vectors $\d$:
\begin{align}
H=-\sum_{\R,\d}\left[t_{\R,\R+\d}^{}a_\R^\dagger b_{\R+\d}^{}+H.c.\right]
\end{align}
The spin-index on the operators shall be suppressed throughout this work. With the Fourier components 
\begin{align}
a_\R^{}&=\frac{1}{\sqrt{N_c}}\sum_\k e^{\ii\k\cdot\R}a_\k^{}\;,\\
\label{FourierB}
b_{\R+\d}^{}&=\frac{1}{\sqrt{N_c}}\sum_\k e^{\ii\k\cdot(\R+\d)}b_\k^{}\;,
\end{align}
this reads for $t_{\R,\R+\d}=t$
\begin{align}
H=-t\sum_{\k}\left[\phi(\k) a_\k^\dagger b_\k^{}+H.c.\right]
\end{align}
with $\phi(\k)=\sum_\d e^{\ii\k\cdot\d}$ the complex structure factor where the sum goes over all nearest-neighbor vectors $\d$. Notice that the phase factor $e^{\ii\k\cdot\d}$ in Eq. (\ref{FourierB}) is important for the definition of the current.\cite{Kotliar03}

We will now define a continuous model by introducing the following Fourier components 
\begin{align}
c_\k&=\frac{1}{\sqrt{A}}\int d^2re^{-\ii\k\cdot\r}c(\r)\;,\\
c(\r)&=\frac{1}{\sqrt{A}}\sum_\k e^{\ii\k\cdot\r}c_\k\;,
\end{align}
where $c=a,b$ and $A$ the area of the sample.

The continuous version of the Hamiltonian thus reads
\begin{align}
\label{Hcont}
H=-t\int d^2r\left[a^\dagger(\r)\phi(-\ii\nabla)b(\r)+H.c.\right]\;.
\end{align}
The gauged Hamiltonian is obtained by replacing $-\ii\nabla\rightarrow -\ii\nabla+\frac{e}{\hbar}\A(\r)$ ($e>0$). Notice that by going back in Fourier space, we obtain the correct Peierls substitution 
\begin{align}
t_{\R,\R+\d}\rightarrow t_{\R,\R+\d}e^{\ii\frac{e}{\hbar}\int_\R^{\R+\d}d\l\A(\l)}
\end{align}
in the case of a gauge field which is constant over one lattice spacing, i.e., for a spatially weakly varying field. Because $e^{\nabla\cdot\d}b(\r)=b(\r+\d)$, we can write Eq. (\ref{Hcont}) as
\begin{align}
H=-t\int d^2r\sum_\d\left[a^\dagger(\r)b(\r+\d)+H.c.\right]\;.
\end{align}
The continuous model thus consists of infinitely many replica of the original lattice model. The unperturbed Hamiltonian is homogeneous, real (not crystalline) momentum is conserved and yet, each particle is bound to hop in the replica where it lives with strict fidelity to the original lattice Hamiltonian. In particular, the lattice anisotropy is fully preserved. Also, the minimal substitution used to include the perturbing vector potential guarantees gauge invariance to all orders.   

For this model, the current can be defined by
\begin{align}
\j(\r)=-\frac{\delta H}{\delta \A(\r)}=\j^P(\r)+\j^D(\r)+\mathcal{O}(A^2)\;,
\end{align}
where the diamagnetic contribution $\j^D$ is linear in the gauge field $\A$.

For the paramagnetic operator, we obtain
\begin{align}
\j^P(\r)=\frac{\ii te}{\hbar}\sum_\d\d\int_0^1ds\left[a^\dagger(\r-s\d)b(\r+(1-s)\d)\right]+H.c.
\end{align}
which consists of a symmetrized version of the paramagnetic current given in Refs. \cite{Scalapino93,Gusynin07} which is obtained from the above formula by setting $s=0$. 

For the diamagnetic contribution, we obtain with (summation over $j$ is implied)
\begin{align}
j^{D,i}(\r)=\int d^2r'\left.\frac{\delta j^i(\r)}{\delta A^j(\r')}\right|_{\A=0}A^j(\r')
\end{align}
the following expression:
\begin{align}
j^{D,i}(\r)&=-\frac{te^2}{\hbar^2}\sum_\d \delta^i\delta^j\int_0^1dsds'\left[a^\dagger(\r-s\d)b(\r+(1-s)\d)\right.\notag\\
&\times\left.\left\{sA^j(\r-ss'\d)+s'A^j(\r+ss'\d)\right\}+H.c.\right]
\end{align}
which again resembles a symmetrized version of the diamagnetic current given in Refs. \cite{Scalapino93,Gusynin07} which this time is obtained from the above formula by setting $s=0$ and $s'=1$. Notice that the diamagnetic current is non-local in the external gauge field. 

In linear response, only ground-state averages enter in the diamagnetic
 current. With the energy per bond per unit area 
\begin{align}
h_{\text{bond}} &= -2 t \langle a^{\dagger}(\r) \; b(\r + \bm \delta)\rangle\;,
\end{align} 
which is independent of both $\r$ and $\d$, the Fourier transform of the paramagnetic and
diamagnetic current are given by: 
\begin{align}
\label{paramagnetic}
&j_\q^{P,i}=\frac{te}{\hbar}\sum_\k \tilde\phi^i(\k,\q)a_\k^\dagger b_{\k+\q}+(\tilde\phi^i(\k,\q))^*b_\k^\dagger a_{\k+\q}\;,\\
&\langle j^{D,i}_{\bm q}\rangle = \chi^{D,i,j}_{\bm q} \;A^{j}_{\bm q}
\end{align}
with 
\begin{align}
&\tilde\phi^i(\k,\q)=\sum_\d\frac{\delta^i}{\q\cdot\d}\left(e^{\ii(\k+\q)\cdot\d}-e^{\ii\k\cdot\d}\right)\;,\\
&\chi^{D,i,j}_{\bm q}= \frac{e^2}{\hbar^2} \; h_{\text{bond}} \sum_{\d} \delta^i \delta^j 
\frac{4}{(\bm q \cdot \d)^2} \; \sin^2(\frac{\bm q \cdot \d}{2}) 
\;.
\end{align}
For $\q\rightarrow0$, we obtain the same expression as in Refs. \cite{Stauber08,Gusynin07}.

The Fourier transform of the particle density of the lattice model is given by $n_\q=\sum_\k(a_\k^\dagger a_{\k+\q}+b_\k^\dagger b_{\k+\q})$.\cite{Bena07} For the charge density $\rho_\q=en_\q$, the continuity equation $\dot\rho_\q-\ii\q\cdot\j_\q=0$ is obeyed for the paramagnetic current operator of Eq. (\ref{paramagnetic}). We can thus consider this operator to be the current operator of the lattice model for general $\q$. In the same manner, the  diamagnetic term is also correct for arbitrary $\q$.

%%%%%%%%%%%%%%%%%%%%%%%%%%%%%%%%%%%%%%%%%%%%%%%%%%%%%%%%%%%%%%%%%%%
% Section : Correlation function
%%%%%%%%%%%%%%%%%%%%%%%%%%%
\section{Correlation function}
We can now determine the current-current correlation function. In terms of the bosonic Matsubara frequencies $\hbar\omega_n=2\pi n/\beta$ ($\beta=1/k_BT$), it is defined by 
\begin{align}
\pi^{i,j}(\q,\ii\omega_n)=\frac{1}{\hbar A}\int_0^{\hbar\beta} d\tau e^{\ii\omega_n\tau}\langle\j_\q^{P,i}(\tau)\j_{-\q}^{P,j}\rangle\;.
\end{align}

We obtain the general expression for the current-current correlation function 
\begin{align}
\label{currentcorrelation}
\pi^{i,j}(\q,\omega)&=\left(\frac{te}{\hbar}\right)^2\frac{-g_s}{(2\pi)^2}\int_{\text{1.BZ}}d^2k\sum_{s,s'=\pm}f_{s\cdot s'}^{i,j}(\k,\q)\nonumber\\
&\times\frac{n_F(E^{s}({\bf k}))-n_F(E^{s'}({\bf k}+{\bf q}))}{E^{s}({\bf k})-E^{s'}({\bf k}+{\bf q})+\hbar\omega+\ii\delta}\;,
\end{align}
with $E^{\pm}({\bf k})=\pm t|\phi(\k)|$ and $n_F(E)=(e^{\beta(E-\mu)}+1)^{-1}$ the Fermi function. 

This is the same expression as for the density-density correlation function of Eq. (\ref{densitycorrelation}), but the band-overlap is now given by
\begin{align}
\label{overlapp}
f_{\pm}^{i,j}(\k,\q)&=\frac{1}{2}\left(\text{Re}\left[\tilde\phi^i(\k,\q)(\tilde\phi^j(\k,\q))^*\right]\right.\\\notag
&\left.\pm\text{Re}\left[\tilde\phi^i(\k,\q)\tilde\phi^j(\k,\q)\frac{\phi^*(\k)}{|\phi(\k)|}\frac{\phi^*(\k+\q)}{|\phi(\k+\q)|}\right]\right)\;.
\end{align}

Due to charge conservation, we have $e^2\omega^2\text{Im}\pi^{0,0}(\q,\omega)=q_iq_j\text{Im}\pi^{i,j}(\q,\omega)$ where summation over double indices is implied. To see this within our notation, we note that $q_i\tilde\phi^i(\k,\q)=\phi(\k)-\phi(\k+\q)$ and thus
\begin{align}
f_\pm^{0,0}(\k,\q)=q^iq^j\frac{f_\pm^{i,j}(\k,\q)}{(|\phi(\k)|\mp|\phi(\k+\q)|)^2}
\end{align}
which proves the relation since $(\hbar\omega/t)^2=(|\phi(\k)|\mp|\phi(\k+\q)|)^2$.

In the Dirac cone approximation, the expressions simplify considerably. Denoting the angle between $\k$ and $\q$ by $\varphi$ and neglecting terms proportional to $\sin\varphi$ which cancel to zero due to the angle integration, we have for the effective band overlap
\begin{widetext}
\begin{align}
f_\pm^{i,i}&=\frac{1}{2}\left(\frac{3a}{2}\right)^2\left(1\pm(-1)^{\delta_{i,y}}\frac{k^2}{q^2}\frac{(q_x^2-q_y^2)}{|\k||\k+\q|}\left[1-2\sin^2\varphi+\frac{q}{k}\cos\varphi\right]\right)\;,\\
f_\pm^{i,j}&=\frac{1}{2}\left(\frac{3a}{2}\right)^2\left(\pm(1-\delta_{i,j})\frac{k^2}{q^2}\frac{2q_xq_y}{|\k||\k+\q|}\left[1-2\sin^2\varphi+\frac{q}{k}\cos\varphi\right]\right)\;,
\end{align}
\end{widetext}
where we introduced the carbon-carbon distance $a=0.14$nm.

The system linearized around the Dirac point is rotationally invariant. We can thus decompose $\pi^{i,j}$ into a longitudinal component $\pi^\parallel$ and transverse component $\pi^\perp$. These are defined by Eq. (\ref{currentcorrelation}) after substitution of the overlap function $f_\pm^{i,j}$ by 
\begin{align}
f_\pm^{\parallel(\perp)}&=\frac{1}{2}\left(\frac{3a}{2}\right)^2\left(1\pm(\mp)\frac{k+q\cos\varphi-2k\sin^2\varphi}{|\k+\q|}\right)\;.
\end{align}
We then recover the general relation
\begin{align}
\pi^{i,j}(\q,\omega)=\frac{q_iq_j}{|\q|^2}\pi^{\parallel}(|\q|,\omega)+(\delta_{i,j}-\frac{q_iq_j}{|\q|^2})\pi^{\perp}(|\q|,\omega)\;.
\end{align}
We note that the overlap function $f_\pm^{0,0}$ in the Dirac approximation is proportional to $f_\pm^{\parallel}$, but with the last term, $2k\sin^2\varphi$, missing.\cite{Wunsch06} 

Due to current conservation and $q^2\pi^\parallel=q_i\pi^{i,j}q_j$, the parallel component of the current-current correlation is related to the density-density correlation by
\begin{align}
q^2\pi^\parallel(|\q|,\omega) =-\langle[\rho_\q,\q\cdot\j_{-\q}]\rangle/(\hbar A)
+e^2\omega^2\pi^{0,0}(|\q|,\omega)\;.
\end{align}
Apart from the constant surface or contact term, which was determined in Ref. \cite{Sabio08} for the linearized Dirac model, we are thus left with the calculation of the perpendicular component $\pi^\perp$ which is related to the magnetic susceptibility $\chi_M(\q,\omega)/\mu_0=\pi^\perp(\q,\omega)/|\q|^2$ for $\omega\ll|\q|$ with $\mu_0$ the magnetic permeability.\cite{Wen}

For the full dispersion, we have $-\langle[\rho_\q,\q\cdot\j_{-\q}]\rangle/(\hbar A)=q_i\chi^{D,i,j}_{\q}q_j$. It is thus often more transparent to deal with the physical response, $\Pi^{i,j}$, which includes the diamagnetic contribution:
\begin{align}
\Pi^{i,j}(\q, \omega) = \pi^{i,j}(\q, \omega) + \chi^{D,i,j}_{\q}
\; \end{align} 
Charge conservation then implies
\begin{align}
q_i \; \Pi^{i,j}(\q, \omega) \; q_j=e^2\omega^2 \pi^{0,0}(\bm q, \omega)
\;.\end{align} 
Notice that the anisotropy of the response for finite $\q$ requires the  full
tensorial structure of $\Pi^{i,j}$. In particular, the  relation between
polarizability and conductivity reads
\begin{align}\label{sigma}
 q_i \; \sigma^{i,j}(\bm q, \omega) \; q_j  = i \omega e^2\pi^{0,0}(\bm q, \omega)
\;.\end{align} 
We will show in the next section that the often used scalar version of Eq. (\ref{sigma}) would not hold for the lattice model even in  the regime where the Dirac approximation is justified.

We finally state the general f-sum rule for a bipartite tight-binding model:
\begin{align}
\frac{2}{\pi}\int_0^{\Lambda_E}d\omega\omega{\rm Im}\pi^{0,0}(\bm q, \omega)=\frac{4h_{\text{bond}}}{\hbar^2} \sum_{\d} \sin^2(\frac{\bm q \cdot \d}{2})
\end{align}
where the energy per bond per unit area of the hexagonal lattice is given by
\begin{align}
h_{\text{bond}} =\frac{g_s}{3A}\sum_\k E^+(\k)\left[n_F(E^{-}({\bf k}))-n_F(E^{+}({\bf k}))\right]\;,
\end{align} 
and the band cutoff $\Lambda_E=6t$.
%%%%%%%%%%%%%%%%%%%%%%%%%%%%%%%%%%%%%%%%%%%%%%%%%%%%%%%%%%%%%%%%%%%
% Section : Results
%%%%%%%%%%%%%%%%%%%%%%%%%%%
\section{Results}
We will now summarize the analytical results obtained for the Dirac approximation  at zero temperature first presented in Ref. \cite{Polini09} and compare them with the numerical results obtained from the hexagonal lattice.

\subsection{Analytical results}
In order to present the analytical results, we express the current-current correlation function of Eq. (\ref{currentcorrelation}), $\pi^\pm(q,\omega)$, by  two dimensionless functions
\begin{align}
\pi^\pm(q,\omega)=\left(\frac{e^2t}{\hbar^2}\right)\left[\pi_0^\pm(q,\omega)+\Delta\pi_\mu^\pm(q,\omega)\right]\;,
\end{align}
where we will use the the superindex $+$ to denote the longitudinal component ($\parallel$) and the superindex $-$ to denote the transverse component ($\perp$).
We restrict the discussion to $\omega\geq0$ since $\pi^\pm(q,-\omega)=\left[\pi^\pm(q,\omega)\right]^*$ and to $\mu\geq0$ due to particle-hole symmetry. $\pi_0^\pm$ contains the contribution for the system at half-filling, i.e., interband contributions, whereas $\Delta\pi_\mu^\pm$ contains the contributions due to the finite chemical potential $\mu$, i.e., intraband contributions. The formulas are given in terms of the Fermi velocity $\hbar v_F=\frac{3}{2}at$.

The results can be written in compact form using two dimensionless, complex functions defined as
\begin{align}
F^\pm(q,\omega)&=\frac{g}{16\pi}\frac{\hbar\omega}{t}\left[1-\left(\frac{v_Fq}{\omega}\right)^2\right]^{\mp\frac{1}{2}}\;,\\
G^\pm(x)&=x\sqrt{x^2-1}\mp \ln\left(x+\sqrt{x^2-1}\right)\;.
\end{align}

Let us first present the results for the undoped system. For large energy cutoff $\Lambda_E\gg1$, we have 
\begin{equation}
\pi_{0}^\pm(q,\omega )=\left[\frac{g}{8\pi}\frac{\Lambda_E}{t}+\ii\pi F^\pm(q,\omega )\right]\,.
\end{equation}%
Notice that the constant cutoff term can be obtained either from the Kramers-Kronig relation or from the continuity equation. This connection gives rise to the so-called f-sum rule.\cite{Sabio08}

The contribution due to the finite chemical potential reads
\begin{align}
\Delta &\pi_\mu^\pm(q,\omega)=\pm\frac{g}{2\pi}\frac{\mu}{t}\frac{\omega^2}{(v_Fq)^2}\mp F^\pm(q,\omega)\left\{
G^\pm\left( x_+\right)\right.\label{eq:DP}\\ 
& \left.-\Theta
\left( x_--1\right) \left[
G^\pm\left(x_-\right) \mp \ii\pi
\right]-\Theta \left(1-x_-\right) G^\pm\left( -x_-\right) \right\} \notag\;
\end{align}
where we defined $x_\pm=\frac{2\mu\pm\hbar\omega}{\hbar v_{F}q}$. 

The above expression for graphene shall be contrasted with the expression for the two-dimensional electron gas. For quadratic dispersion $\epsilon_\k=\hbar^2\k^2/(2m)$, we have
\begin{align}
\pi^\pm(q,\omega)=\left(\frac{e}{\hbar}\right)^2\left\{\frac{\mu}{2\pi}\pm\frac{\omega^2}{q^2}\frac{m}{2\pi}\left(1-\left[1-\left(\frac{v_Fq}{\omega}\right)^2\right]^{\mp\frac{1}{2}}\right)\right\}\;,
\end{align}
where the term proportional to $\mu=\epsilon_{k_F}$ corresponds to the contact term which is canceled by the diamagnetic contribution.

Eq. (\ref{eq:DP}) can be written as real and imaginary part in terms of three real dimensionless functions
\begin{align}
\label{eq:realG}
f^\pm(q,\omega )& =\frac{g}{16\pi}\frac{\hbar\omega}{t}\left|1-\left(\frac{v_Fq}{\omega}\right)^{2}\right|^{\mp\frac{1}{2}}\,,\\
G_{>}^\pm(x)& =x\sqrt{x^{2}-1}\mp\cosh ^{-1}(x)\,, \quad x>1 \, , \notag \\
G_{<}^\pm(x)& =\pm x\sqrt{1-x^{2}}-\cos ^{-1}(x)\, , \quad |x|<1\, .\notag
\end{align}
The lengthy expressions are given in the appendix.

Let us now discuss two limiting cases. For the long wavelength limit $q\rightarrow0$, we obtain 
\begin{align}
\pi^\pm(q\rightarrow0,\omega)&=\frac{e^2}{\hbar}\frac{g}{8\pi}\omega\left[-\frac{2\mu}{\hbar\omega}+\ln\left|\frac{2\mu+\hbar\omega}{2\mu-\hbar\omega}\right|\right.\nonumber\\\label{HighEnergieAppr}
&+\left.\ii\frac{\pi}{2}\Theta(\hbar\omega-2\mu)\right]\;.
\end{align}
Using the RPA-approximation for the longitudinal part, the above expansion leads to plasmon excitations for which the logarithmic term is usually neglected.\cite{Wunsch06,Hwang07} Due to the sign change of the photon propagator in the case of transverse modes, the denominator of the RPA-approximation cannot become zero for the perpendicular part without the logarithmic term. But including it leads to a new transverse electromagnetic mode in graphene.\cite{Ziegler07}

For the static case, we obtain the following formula which was first given in Ref. \cite{Ando09}:
\begin{align}
\pi^-(q,\omega=0)=\frac{e^2}{\hbar}\frac{g}{8\pi}v_Fq\Theta(q-2k_F)G_<^-(\frac{2k_F}{q})
\end{align}
where $k_F=\mu/(\hbar v_F)$. The parallel component $\pi^+$ is zero. For fixed $q$, $\pi_\mu^-$  is only non-zero for $\mu<\hbar v_Fq/2$ and since $\int_0^1dxG_<^-(x)=4/3$, the limit $q\rightarrow0$ leads to the well known delta function for the diamagnetic susceptibility of graphene: 
\begin{align}
\label{diadelta}
\chi_M=-\mu_0\frac{g}{6\pi}e^2v_F^2\delta(\mu)
\end{align}

%%%%%%%%%%%%%%%%%%%%%%%%%%%%%%%%%%%%%%%%%%%%%%%%%%%%%%%%%%%%%%%%%%%
% Section : Numerical Results
%%%%%%%%%%%%%%%%%%%%%%%%%%%
\subsection{Numerical Results}
\begin{figure}[t]
\begin{center}
\includegraphics[angle=0,width=0.8\linewidth]{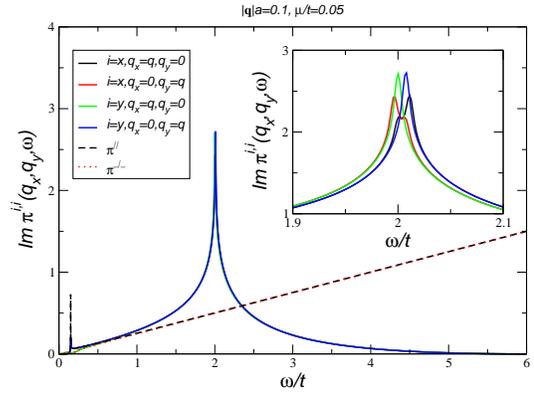}
\caption{The imaginary part of the current-current correlation function $\text{Im} \pi^{i,i}(q_x,q_y,\omega)$ as function of the energy $\omega$ at $k_BT/t=0.01$ for different directions with $|\q|a=0.1$. The results obtained from the Dirac-cone approximation $\pi^\parallel(q,\omega)$ and $\pi^\perp(q,\omega)$ are also shown (dashed lines).} 
  \label{fig:ImP}
\end{center}
\end{figure}

We shall now compare the analytical results of the linearized, isotropic Dirac model with the numerical results obtained from the hexagonal lattice. In Fig. \ref{fig:ImP}, we show the imaginary part of the current-current correlation function $\text{Im} \pi^{i,i}(q_x,q_y,\omega)$ as function of the energy $\hbar\omega$ at $k_BT/t=0.01$ for different directions with $|\q|a=0.1$. The results obtained from the Dirac-cone approximation $\pi^\parallel(q,\omega)$ and $\pi^\perp(q,\omega)$ are also shown (dashed lines). Clearly, there are strong differences for energies $\hbar\omega>t$ due to the van Hove singularity. The inset shows that there is a peak splitting for the different directions due to the different contributions of the three $M$-points, also present in the charge response.\cite{Stauber10}

\begin{figure}[t]
\begin{center}
\includegraphics[angle=0,width=0.8\linewidth]{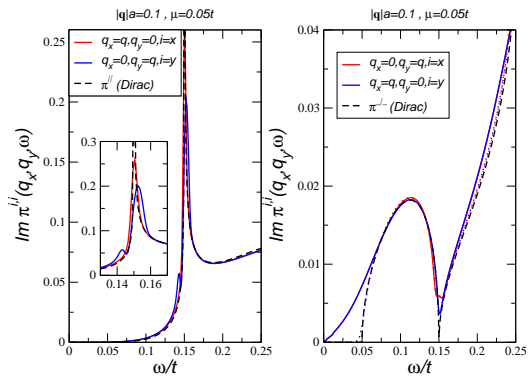}
\caption{The imaginary part of the current-current correlation function $\text{Im} \pi^{i,i}(q_x,q_y,\omega)$ as function of the energy $\omega$ at $k_BT/t=0.01$ for different directions with $|\q|a=0.1$. Left hand side: The wave vector $\q$ is parallel to the current. Right hand side: The wave vector $\q$ is perpendicular to the current.The results obtained from the Dirac-cone approximation $\pi^\parallel(q,\omega)$ and $\pi^\perp(q,\omega)$ are also shown, respectively (dashed lines). Inset: Energy region around the resonant energy $\omega=v_Fq$. Right hand side: Also the curves for lower temperature $k_BT/t=0.001$ are shown (dotted lines).} 
  \label{fig:ImPle}
\end{center}
\end{figure}

In Fig. \ref{fig:ImPle}, the same curves are shown, but for lower energies. On the left hand side, the wave vector $\q$ is parallel to the current and on the right hand side perpendicular. The results obtained from the Dirac-cone approximation $\pi^\parallel(q,\omega)$ and $\pi^\perp(q,\omega)$ are also shown (dashed lines). 

For the perpendicular contribution of $\pi^{i,i}$ (right hand side), clear differences are seen for lower energies due to the finite temperature $k_BT/t=0.01$ used in the numerical calculation. This results in a thermal broadening of the delta-function of Eq. (\ref{diadelta}) and is responsible for the diamagnetism found in graphene\cite{Geim10} since intrinsic doping leads to $\mu\neq0$. The dotted lines on the right hand side refer to the same curves but at lower temperature $k_BT/t=0.001$ which agrees well with the Dirac cone approximation now also at low energies.

When $\q$ is in $x$-direction which was chosen to be the high symmetry axis which connects the $\Gamma$- and $M$-point of the Brillouin zone, there is good agreement with the result coming from the Dirac cone approximation (except for the deviations in $\pi^\perp$ due to temperature, mentioned before). When $\q$ is in $y$-direction, we observe a peak splitting around the resonant energy $\hbar\omega=\hbar v_Fq$ (see inset on the left hand side). There are thus lattice effects which show up even in the regime where the Dirac cone approximation and where the system should be isotropically invariant.

%%%%%%%%%%%%%%%%%%%%%%%%%%%%%%%%%%%%%%%%%%%%%%%%%%%%%%%%%%%%%%%%%%%
% Section : Summary and Conclusions
%%%%%%%%%%%%%%%%%%%%%%%%%%%
\section{Summary and Conclusions}
We have discussed the dynamical current-current correlation function of the hexagonal lattice and of graphene modeled by the linearized Dirac model. To define a local current operator, we introduced a continuum-replica of the original lattice model. The resulting paramagnetic current operator obeys the continuity equation with respect to the density operator defined on the original lattice. The diamagnetic response is non-local.

We then gave explicit expressions of the current-current correlation function for the honeycomb lattice and defined the longitudinal and transverse component in case of the rotationally invariant Dirac model. For the Dirac model, explicit analytical expressions were given where the results for the longitudinal component can be obtained via the continuity equation from the density-density correlation function, as was discussed in detail.

In the last part of this paper, we showed that in the honeycomb lattice, the longitudinal and transverse component cannot be defined for energies around the resonant energy $\hbar\omega=\hbar v_Fq$. This is reminiscent to the fact that also the polarizability is not well described by the Dirac approximation for these energies.\cite{Stauber10} The scalar relation between the conductivity and the polarizability which makes use of the fact that there is a parallel component does thus not hold for the lattice model. This might be important for first principle studies which make use of this relation.

\section{Acknowledgments}
This work has been supported by grants PTDC/FIS/101434/2008 and FIS2010-21883-C02-02.
\begin{figure}[t]
\begin{center}
\includegraphics[angle=0,width=0.5\linewidth]{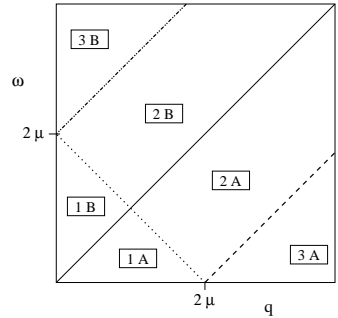}
\end{center}
\caption{Display of the different regions characterizing the current-current correlation function given in the appendix. The regions are limited by straight lines $\protect\omega=q$ (solid), $%
\protect\omega=q-2\protect\mu$ (dashed) and $\protect\omega=2\protect\mu-q$
(dotted) where we set $\hbar=v_F=1$.}
\label{fig:Regions}
\end{figure}

%%%%%%%%%%%%%%%%%%%%%%%%%%%%%%%%%%%%%%%%%%%%%%%%%%%%%%%%%%%%%%%%%%%
% Section : Appendix
%%%%%%%%%%%%%%%%%%%%%%%%%%%
\section{appendix}
Here, we shall present the real and imaginary part of $\pi^\pm$ in terms of the three real dimensionless functions 
\begin{align}
\label{eq:realGApp}
f^\pm(q,\omega )& =\frac{g}{16\pi}\frac{\hbar\omega}{t}\left|1-\left(\frac{v_Fq}{\omega}\right)^{2}\right|^{\mp\frac{1}{2}}\,,\\
G_{>}^\pm(x)& =x\sqrt{x^{2}-1}\mp\cosh ^{-1}(x)\,, \quad x>1 \, , \notag \\
G_{<}^\pm(x)& =\pm x\sqrt{1-x^{2}}-\cos ^{-1}(x)\, , \quad |x|<1\, .\notag
\end{align}

 For the imaginary part, the additional terms at finite doping then read in the language of Fig.~\ref{fig:Regions}:
\begin{widetext}
\begin{align}
\text{Im}\Delta \pi_\mu^\pm(q,\omega )=-f^\pm(q,\omega )\times\left\{
\begin{array}{ll}
G_{>}^\pm(\frac{2\mu -\hbar\omega }{\hbar v_Fq})-G_{>}^\pm(\frac{2\mu +\hbar\omega }{\hbar v_Fq}) & ,\text{ 1 A}
\\[1.5ex]
\pi & ,\text{ 1 B} \\
-G_{>}^\pm(\frac{2\mu +\hbar\omega }{\hbar v_Fq}) & ,\text{ 2 A} \\
-G_{<}^\pm(\frac{\hbar \omega -2\mu }{\hbar v_Fq}) & ,\text{ 2 B} \\
0 & ,\text{ 3 A} \\
0 & ,\text{ 3 B}
\end{array}%
\right. 
\end{align}

For the real part, we get in the language of Fig.~\ref{fig:Regions}:
\begin{align}
\text{Re}& \Delta \pi_\mu^\pm(q,\omega )=\pm\frac{g\mu }{2\pi t}\frac{\omega^2}{(v_Fq)^2}
\mp f^\pm(q,\omega )\times\left\{
\begin{array}{ll}
\pi & ,\text{ 1 A} \\[1.3ex]
-G_{>}^\pm(\frac{2\mu -\hbar\omega }{\hbar v_Fq})+G_{>}^\pm(\frac{2\mu +\hbar\omega }{\hbar v_Fq}) & ,\text{ 1 B}
\\[1.3ex]
-G_{<}^\pm(\frac{\hbar\omega -2\mu }{\hbar v_Fq}) & ,\text{ 2 A} \\[1.3ex]
G_{>}^\pm(\frac{2\mu +\hbar\omega }{\hbar v_Fq}) & ,\text{ 2 B} \\[1.3ex]
- G_{<}^\pm(\frac{\hbar\omega -2\mu }{\hbar v_Fq})+ G_{<}^\pm(\frac{2\mu +\hbar\omega }{\hbar v_Fq}) & ,\text{ 3 A}
\\[1.3ex]
G_{>}^\pm(\frac{2\mu +\hbar\omega }{\hbar v_Fq})-G_{>}^\pm(\frac{\hbar\omega -2\mu }{\hbar v_Fq}) & ,\text{ 3 B}
\end{array}%
\right. 
\end{align}
\end{widetext}
Since 
\begin{align}
G^\pm(x)=\left\{
\begin{array}{ll}
G_{>}^\pm(x)& , x>1 \\
\pm \ii G_{<}^\pm(x)& ,|x|<1
\end{array}%
\right.  
\end{align}
this agrees with the complex expression given in Eq. (\ref{eq:DP}). For more details, see Ref. \cite{Wunsch06} and Ref. \cite{Polini09}.

%%%%%%%%%%%%%%%%%%%%%%%%%%%%%%%%%%%%%%%%%%%%%%%%%%%%%%%%%%%%%%%%%%%
% Section : Bibliography
%%%%%%%%%%%%%%%%%%%%%%%%%%%


\begin{thebibliography}{99}
\bibitem{footnote}
Note that the functional form of the band-overlap function depends on the phase of the structure factor.
\end{thebibliography}
\end{document}